\documentclass[12pt]{article}
\usepackage{caption}
\usepackage{amsmath}
\usepackage{amstext}
\usepackage{amsthm}
\usepackage{amssymb}
\usepackage{amsopn}
\usepackage{mathrsfs} 
\usepackage{dsfont} 
\usepackage[bottom]{footmisc}
\usepackage{indentfirst}
\usepackage{upref}
\usepackage{cite}
\usepackage{units}
\usepackage{fancybox}
\usepackage{graphicx}
\usepackage
{floatflt}
\numberwithin{equation}{section}

\renewcommand{\baselinestretch}{1.3}
\newcommand{\ddx}{\textrm{d}^d x\:}
\newcommand{\dz}{\textrm{d} z\:}
\newcommand{\gYM}{g_{\textrm{YM}}}
\begin{document}
\begin{titlepage}
\renewcommand{\baselinestretch}{1.1}
\title{\begin{flushright}
\normalsize{MZ-TH/09-39}
\bigskip
\end{flushright}
Running Gauge Coupling in\\
Asymptotically Safe Quantum Gravity}
\date{}
\author{J.-E. Daum, U. Harst and M. Reuter\\
{\small Institute of Physics, University of Mainz}\\[-0.2cm]
{\small Staudingerweg 7, D-55099 Mainz, Germany}}
\maketitle
\begin{abstract} 
We investigate the non-perturbative renormalization group behavior of the \linebreak gauge coupling constant using a truncated form of the functional flow equation for the effective average action of the Yang-Mills--gravity system. We find a non-zero quantum gravity correction to the standard Yang-Mills beta function which has the same sign as the gauge boson contribution. Our results fit into the picture according to which Quantum Einstein Gravity (QEG) is asymptotically safe, with a vanishing gauge coupling constant at the non-trivial fixed point.
\end{abstract}
\end{titlepage}
\newpage 
\section{Introduction}
Recently a lot of efforts went into the computation of gravitational corrections to the beta function of the running Yang-Mills coupling constant. Robinson and Wilczek \cite{ymg-robwil, ymg-robthesis}, in an effective field theory setting, obtained a non-zero correction at the one-loop level. It has the same negative sign as the familiar term already present in absence of gravity, and so it would render even pure abelian theories asymptotically free. After Pietrykowski \cite{ymg-piet} had realized that this result is gauge fixing dependent, Toms \cite{ymg-toms} reanalyzed the problem using the Vilkovisky-DeWitt method. In a manifestly gauge invariant as well as gauge fixing independent formulation of the effective action he finds that the quantum gravity contributions to the running charge vanish.

All of these computations employ the dimensional regularization scheme. In \cite{ymg-ebert}, Ebert, Plefka and Rodigast pointed out that its use might be problematic since it is insensitive to quadratic divergences, and it is precisely such quadratic divergences that are responsible for the non-zero result obtained in \cite{ymg-robwil, ymg-robthesis}. Using a cutoff regularization instead they found that all gravitational quadratic divergences cancel so that there is again no correction to the beta function. Thereafter Tang and Wu \cite{ymg-tangwu} argued that the use of a cutoff regularization is not permissible here since it does not respect gauge invariance. Performing a calculation in a scheme which both retains quadratic divergences and preserves gauge invariance (``loop regularization'') they obtained a non-zero gravitational correction to the one-loop beta function. Furthermore, Toms \cite{ymg-toms2} demonstrated that, with a cosmological constant included, also dimensional regularization yields a non-vanishing gravitational correction, albeit of a different type.

In the present paper we shall analyze the running of the gauge coupling constant in the framework of the Asymptotic Safety approach to quantum gravity \cite{wein,wein-kall,wein-bern,mr,percadou,oliver1,frank1,oliver2,oliver3, oliver4, souma,frank2, prop, oliverbook, perper1, codello, yukawa, litimgrav, frankmach, BMS, gies-ghost, oliverfrac, jan1, jan2, creh1,creh2,crehroberto, elisa1, brightontalk, elisa2,je1, max-pert,max,livrev}. Contrary to the calculations mentioned above it considers metric gravity not merely an effective but rather a fundamental quantum field theory, with the continuum limit taken at a non-trivial renormalization group (RG) fixed point \cite{wein, wein-kall, wein-bern}. Instead of perturbation theory, the main tool will be the gravitational average action \cite{mr} and a suitably truncated form of the associated functional RG equation (FRGE). Originally developed for matter field theories \cite{avact,nonabavact,ym,mrcwqcd,giesymfp,avactrev} the effective average action turned out an ideal tool for investigating the RG flow of Quantum Einstein Gravity (QEG)\cite{mr,percadou,oliver1,frank1,oliver2,oliver3, oliver4, souma,frank2, prop, oliverbook, perper1, codello, yukawa, litimgrav, frankmach, BMS, gies-ghost, oliverfrac, jan1, jan2, creh1,creh2,crehroberto, elisa1, brightontalk, elisa2,je1} and exploring its potential physics implications \cite{bh,erick1,cosmo1,cosmo2,entropy,esposito,h1,h2,h3,girelli,lhc,mof}.

The purpose of this paper is twofold: First, we develop a general framework for both exact and approximate (``truncated'') investigations of the Yang-Mills--gravity system by means of a {\it gauge invariant} running effective action. In particular we shall see that because of the semi-direct product structure of the pertinent gauge group there arises a subtlety as for the appropriate construction of the ghost action. Second, we use the resulting framework in order to find the RG flow in a simple truncation of the space of actions which, however, is general enough to allow for an approximate determination of the beta function of the Yang-Mills coupling constant in presence of quantized gravity.

The remaining sections of this paper are organized as follows. In Section 2 we describe the general setup of the effective average action for a Yang-Mills field coupled to gravity, the corresponding FRGE in particular. Then, in Section 3, we discuss the problem of background covariant ghost actions and explain its solution. The explicit computation of the running gauge coupling is performed in Section 4, and Section 5 contains a summary as well as a brief discussion of various immediate applications of our results.
\section{The Functional RG Setup}
Schematically the dynamics of the Yang-Mills--gravity system is governed by the path integral
\begin{eqnarray}\label{PI-sketchy}
{\cal Z} = \int\:{\cal D}\gamma_{\mu\nu}{\cal D}{\cal A}^a_\mu\:{\rm e}^{-S[\gamma, {\cal A}]}
\end{eqnarray}

\noindent Here $\gamma_{\mu\nu}$ and ${\cal A}^a_\mu$ are the quantum metric and the quantum gauge field, respectively, and $S$ denotes the bare action. As usual, both of these fields are supposed to transform tensorially with respect to diffeomorphisms, $\delta_{\rm D}$. In addition, ${\cal A}^a_\mu$ defines a connection with respect to Yang-Mills gauge transformations, $\delta_{\rm YM}$. Denoting the vector field that generates the diffeomorphism by $v^\mu$ and the parameter of the Yang-Mills transformation by $\lambda^a$, we have (${\cal L}_v$ denotes the Lie derivative along $v^\mu$):
\begin{eqnarray}
\label{full-quantum-metric-diff}\delta_{\rm D}(v) \gamma_{\mu\nu} &=& {\cal L}_v \gamma_{\mu\nu}\\ 
\label{full-quantum-gaugefield-diff}\delta_{\rm D}(v) {\cal A}^a_\mu &=& {\cal L}_v {\cal A}^a_\mu\\
\label{full-quantum-gaugefield-ym}\delta_{\rm YM}(\lambda) {\cal A}^a_\mu &=& - \partial_\mu \lambda^a + f^{abc}\lambda^b {\cal A}^c_\mu
\end{eqnarray}

\noindent From now on we will assume the Yang-Mills gauge group to be $SU(N)$, so $a$ runs from $1$ to $N^2 - 1$, $f^{abc}$ are the associated structure constants. We demand $S$ to be invariant under both $\delta_{\rm D}$ and $\delta_{\rm YM}$.

Employing the background formalism, the dynamical fields are decomposed according to
\begin{eqnarray}\label{decomp}
\gamma_{\mu\nu} \equiv \bar{g}_{\mu\nu} + h_{\mu\nu} \hspace{0.6cm}\mbox{and}\hspace{0.6cm} {\cal A}^a_\mu \equiv \bar{A}^a_\mu + a^a_\mu
\end{eqnarray}

\noindent with fixed, but arbitrary background configurations $\bar{g}_{\mu\nu}$ and $\bar{A}^a_\mu$ and fluctuations $h_{\mu\nu}$ and $a^a_\mu$ \cite{back}. Assuming a translational invariant measure, the fluctuation fields will replace the full quantum fields as the variables of integration in (\ref{PI-sketchy}).

There are now two possibilities to realize the gauge transformations $\delta_{\rm D}$ and $\delta_{\rm YM}$ at the level of the background decomposition:
\begin{itemize}
\item The {\it background gauge transformations} $\delta^{\rm B}_{\rm D}$ and $\delta^{\rm B}_{\rm YM}$ are defined such that under diffeomorphisms all the fields transform tensorially, i.\:e. 
\begin{eqnarray}\label{background-diff}
\delta^{\rm B}_{\rm D}(v) \Phi &=& {\cal L}_v \Phi\:,\hspace{0.2cm} \Phi \in\{\bar{g}_{\mu\nu}, h_{\mu\nu}, \bar{A}^a_\mu, a^a_\mu\}
\end{eqnarray}
\noindent With respect to Yang-Mills transformations, the background gauge field transforms as a connection and the fluctuation transforms homogeneously:
\begin{eqnarray}
\label{background-YM-gaugefield} \delta^{\rm B}_{\rm YM}(\lambda) \bar{A}^a_\mu &=& - \partial_\mu \lambda^a + f^{abc}\lambda^b \bar{A}^c_\mu, \hspace{1cm} \delta^{\rm B}_{\rm YM}(\lambda) a^a_\mu = f^{abc}\lambda^b a^c_\mu
\end{eqnarray} 
\item On the other hand, we can define {\it true gauge transformations} $\delta^{\rm G}_{\rm D}$ and $\delta^{\rm G}_{\rm YM}$ by requiring that these shall only affect the fluctuations, not the background fields:
\begin{eqnarray}
\label{true-diff-metric} \delta^{\rm G}_{\rm D}(v) \bar{g}_{\mu\nu} &=& 0, \hspace{0.6cm} \delta^{\rm G}_{\rm D}(v) h_{\mu\nu} = {\cal L}_v \big(\bar{g}_{\mu\nu} + h_{\mu\nu}\big)\\
\label{true-diff-gaugefield} \delta^{\rm G}_{\rm D}(v) \bar{A}^a_\mu &=& 0, \hspace{0.6cm} \delta^{\rm G}_{\rm D}(v) a^a_\mu = {\cal L}_v \big(\bar{A}^a_\mu + a^a_\mu\big)\\
\label{true-ym-gaugefield} \delta^{\rm G}_{\rm YM}(\lambda) \bar{A}^a_\mu &=& 0, \hspace{0.6cm} \delta^{\rm G}_{\rm YM}(\lambda) a^a_\mu = - \partial_\mu \lambda^a + f^{abc}\lambda^b \big(\bar{A}^c_\mu + a^c_\mu\big)
\end{eqnarray}
\end{itemize}

The crucial idea is to choose a gauge fixing term $S^{\rm gf}$ that breaks only the true gauge invariance but retains background gauge invariance. It gives rise to an associated ghost action $S_{\rm gh}$ in the usual way. Introducing ghost fields ${\cal C}_\mu$ and $\bar{{\cal C}}^\mu$ for the diffeomorphisms and $\Sigma^a$ and $\bar{\Sigma}^a$ for the Yang-Mills transformations, respectively, we arrive at the following path integral which depends parametrically on the background fields: 
\begin{eqnarray}\label{PI-more-concrete}
{\cal Z} = \int\:{\cal D}h_{\mu\nu}{\cal D}a^a_\mu{\cal D}{\cal C}_\mu{\cal D}\bar{{\cal C}}^\mu{\cal D}\Sigma^a{\cal D}\bar{\Sigma}^a\:{\rm e}^{-S[\bar{g}+h, \bar{A}+a] - S^{\rm gf} - S_{\rm gh}}
\end{eqnarray}

In order to set up a functional RG equation, we follow the well-known construction of the effective average action \cite{avact, nonabavact, avactrev} and add a higher derivative IR cutoff term $\Delta_k S$ that is quadratic in the fluctuations:
\begin{eqnarray}\label{IR-cutoff}
\Delta_k S &=& \frac{1}{2} \kappa^2 \int\:{\rm d}^d x\:\sqrt{\bar{g}}\:h_{\mu\nu}{{\cal R}^{\rm grav}_k [\bar{g}]}^{\mu\nu\rho\sigma}h_{\rho\sigma}\nonumber\\
&& + \frac{1}{2}\int\:{\rm d}^d x\:\sqrt{\bar{g}}\:a^a_\mu{{\cal R}^{\rm YM}_k [\bar{g}, \bar{A}]}^{a\mu b\nu}a^b_\nu\nonumber\\
&& + \sqrt{2} \int\:{\rm d}^d x\:\sqrt{\bar{g}}\:\big(\bar{\cal{C}}, \bar{\Sigma}\big){{\cal R}^{\rm gh}_k [\bar{g}, \bar{A}]}\bigg(\!\!\begin{array}{c} {\cal C} \\[-0.2cm] \Sigma \end{array}\!\! \bigg)
\end{eqnarray}

\noindent with $\kappa \equiv (32 \pi \hat{G})^{-\frac{1}{2}}$ and $\hat{G}$ denoting Newton's constant (see below). Furthermore, adding appropriate source terms enables us to easily compute expectation values of the quantum fields. With the notation
\begin{eqnarray}
\label{gf-vev} A^a_\mu &\equiv& \langle {\cal A}^a_\mu \rangle, \hspace{0.6cm} \bar{a}^a_\mu \equiv \langle a^a_\mu \rangle \\
\label{matric-vev} g_{\mu\nu} &\equiv& \langle \gamma_{\mu\nu} \rangle, \hspace{0.6cm} \bar{h}_{\mu\nu} \equiv \langle h_{\mu\nu} \rangle \\
\label{diff-gh-vev} \xi^\mu &\equiv& \langle {\cal C}^\mu \rangle, \hspace{0.6cm} \bar{\xi}_\mu \equiv \langle \bar{{\cal C}}_\mu \rangle \\
\label{ym-gh-vev} \Upsilon^a &\equiv& \langle \Sigma^a \rangle, \hspace{0.6cm} \bar{\Upsilon}^a \equiv \langle \bar{\Sigma}^a \rangle 
\end{eqnarray}

\noindent the background decomposition (\ref{decomp}) now reads
\begin{eqnarray}\label{decomp-vev}
g_{\mu\nu} \equiv \bar{g}_{\mu\nu} + \bar{h}_{\mu\nu} \hspace{0.6cm}\mbox{and}\hspace{0.6cm} A^a_\mu \equiv \bar{A}^a_\mu + \bar{a}^a_\mu
\end{eqnarray}

\noindent The Legendre transformation of the now $k$-dependent functional ${\rm ln}\:{\cal Z}_k$ with respect to the sources leads to the effective average action then \cite{avact,nonabavact,avactrev}:
\begin{eqnarray}\label{eaa}
\Gamma_k[\bar{h}_{\mu\nu}, \bar{a}^a_\mu, \xi^\mu, \bar{\xi}_\mu, \Upsilon^a, \bar{\Upsilon}^a; \bar{g}_{\mu\nu}, \bar{A}^a_\mu] \equiv \Gamma_k[g_{\mu\nu}, \bar{g}_{\mu\nu}, A^a_\mu, \bar{A}^a_\mu, \xi^\mu, \bar{\xi}_\mu, \Upsilon^a, \bar{\Upsilon}^a]
\end{eqnarray}

\noindent Its scale dependence is governed by the functional renormalization group equation (FRGE)
\begin{eqnarray}\label{FRGE}
\partial_t \Gamma_k = \frac{1}{2}{\rm STr} \Big[\Big(\Gamma_k^{(2)} + {\cal R}_k\big(\Delta\big)\Big)^{-1}\big(\partial_t{\cal R}_k\big(\Delta\big)\Big)\Big]
\end{eqnarray}

\noindent where $\Gamma_k^{(2)}$ denotes the Hessian of $\Gamma_k$ with respect to all fluctuation and ghost expectation values, $\Delta$ is some suitably chosen generalized Laplacian, and $t \equiv {\rm ln}\:k$ is the ``RG time''.

Since we deal with two gauge invariances, a remark concerning the notation is in order: We write $D \equiv \partial + \Gamma$ and $\nabla \equiv \partial + A$ for the covariant derivatives that are constructed by means of $\Gamma^\rho_{\mu\nu} = \frac{1}{2}g^{\rho\sigma}\big(\partial_\nu g_{\sigma\mu} + \partial_\mu g_{\sigma\nu} - \partial_\sigma g_{\mu\nu}\big)$ and $A^a_\mu$, respectively; the covariant derivative containing both of these connections is denoted by ${\cal D} \equiv \partial + A +\Gamma$. By adding a bar, we denote their analogues evaluated on the background configurations.

\section{On the Construction of \\Background Gauge Invariant Ghost Actions}
\subsection{Motivation}
Since we have to fix two gauge invariances by two gauge conditions, $\textrm{F}_\mu(h_{\mu\nu}; \bar{g}_{\mu\nu}, \bar{A}^a_\mu)$ and $\textrm{G}^a(a^a_\mu; \bar{g}_{\mu\nu}, \bar{A}^a_\mu)$ for the diffeomorphisms and the $SU(N)$ transformations, respectively, the associated Faddeev-Popov operator will in general consist of four components. (Here we have already assumed that the diffeomorphism and the $SU(N)$ gauge condition only involve the metric and the gauge field fluctuation separately.) The corresponding classical ghost action will then be of the form
\begin{equation}
\begin{aligned}
S_{\rm gh}[h, a, {\cal C}, &\bar{\cal C}, \Sigma, \bar{\Sigma}; \bar{g}, \bar{A}] = - \int\:{\rm d}^d x\:\sqrt{\bar{g}}\:\Big(\kappa^{-1}\:\bar{{\cal C}}_\mu \bar{g}^{\mu\nu}\frac{\partial {\rm F}_\nu}{\partial h_{\rho\sigma}}\delta_{\rm D}^{\rm G}({\cal C}) h_{\rho\sigma}+ \\
& + \kappa^{-1}\:\bar{{\cal C}}_\mu \bar{g}^{\mu\nu}\frac{\partial {\rm F}_\nu}{\partial h_{\rho\sigma}}\delta_{\rm YM}^{\rm G}(\Sigma) h_{\rho\sigma} + \hat{g}\: \bar{\Sigma}^a \frac{\partial {\rm G}^a}{\partial a^b_\mu}\delta_{\rm D}^{\rm G}({\cal C}) a^b_\mu + \hat g \: \bar{\Sigma}^a \frac{\partial {\rm G}^a}{\partial a^b_\mu}\delta_{\rm YM}^{\rm G}(\Sigma) a^b_\mu \Big) 
\end{aligned}
\label{Geistwirk-allg}
\end{equation}

Since we are going to neglect renormalization effects in the ghost sector, the evolution equation for $\Gamma_k$ will contain only the classical ghost action but with the quantum ghost fields replaced by their vacuum expectation values, and the full classical fields $g_{\mu\nu}$ and $A^a_\mu$ identified with their background configurations $\bar{g}_{\mu\nu}$ and $\bar{A}^a_\mu$, respectively; stated differently, in this class of approximations the fluctuations $\bar{h}_{\mu\nu}$ and $\bar{a}^a_\mu$ can be set to zero in $S_{\rm gh}$ even before $S^{(2)}_{\rm gh}$ is computed.

Looking at the $\bar{\Upsilon}$-$\xi$ part of the resulting ghost action we encounter a serious problem: Employing ${\rm G}^a (a; \bar{g}, \bar{A}) \equiv \bar{{\cal D}}^\mu a^a_\mu$ as gauge condition, it is given by
\begin{equation} \label{ups-xi-geist-alt}
 \left(S_{\rm gh}\:[\xi, \bar{\Upsilon}; \bar{g}, \bar{A}]\right)_{\bar{\Upsilon}\xi} = - \int\:{\rm d}^d x\:\sqrt{\bar{g}}\:\big[\bar{\Upsilon}^a \bar{{\cal D}}^{\nu\:ab} \big(\xi^\rho \partial_\rho \bar{A}^b_\nu + (\partial_\nu \xi^\rho) \bar{A}^b_\rho\big)\big]
\end{equation}

\noindent Here the covariant background derivative $\bar{{\cal D}}_\mu$ acts on the ordinary Lie derivative of an $SU(N)$ background connection; therefore, this part of the ghost action is {\it not} $\delta^{\rm B}_{\rm YM}$-invariant. In order to resolve this issue, it is useful to look at it from a more abstract point of view.
\subsection{Ward operators and their algebra}
To begin with, we consider an arbitrary functional $F[\gamma_{\mu\nu}, {\cal A}^a_\mu, {\cal C}^\mu, \bar{{\cal C}}_\mu, \Sigma^a, \bar{\Sigma}^a]$ of the dynamical fields at hand. At this stage the splitting into fluctuations and background configurations has not yet been performed. An infinitesimal gauge transformation of $F$, considered a scalar functional of its arguments, consists of a diffeomorphism along $v^\mu$ and an $SU(N)$ transformation with parameters $\lambda^a$. It can be implemented as
\begin{equation}
\begin{aligned}
&F[\gamma + \delta_{\rm D} (v)\gamma + \delta_{\rm YM} (\lambda)\gamma , {\cal A} + \delta_{\rm D} (v){\cal A} + \delta_{\rm YM} (\lambda){\cal A}, {\cal C} + \delta_{\rm D} (v){\cal C} + \delta_{\rm YM} (\lambda){\cal C},\\
&\hspace{0.7cm}\bar{{\cal C}} + \delta_{\rm D} (v)\bar{{\cal C}} + \delta_{\rm YM} (\lambda)\bar{{\cal C}}, \Sigma + \delta_{\rm D} (v){\Sigma} + \delta_{\rm YM} (\lambda){\Sigma}, \bar{\Sigma} + \delta_{\rm D} (v)\bar{\Sigma} + \delta_{\rm YM} (\lambda)\bar{\Sigma}] \\
&= F[\gamma, {\cal A}, {\cal C}, \bar{{\cal C}}, \Sigma, \bar{\Sigma}] - {\cal W}_{\rm D}(v)F[\gamma, {\cal A}, {\cal C}, \bar{{\cal C}}, \Sigma, \bar{\Sigma}] - {\cal W}_{\rm YM}(\lambda)F[\gamma, {\cal A}, {\cal C}, \bar{{\cal C}}, \Sigma, \bar{\Sigma}]\\
&\hspace{0.5cm} + {\cal O} (v^2, v \lambda, \lambda^2)
\end{aligned}
\end{equation}

\noindent with the corresponding Ward operators generating diffeomorphisms, 
\begin{equation}
\begin{aligned}
{\cal W}_{\rm D}(v) &\equiv -\int\:{\rm d}^d x\:\bigg(\delta_{\rm D}(v)\gamma_{\mu\nu} (x)\frac{\delta}{\delta \gamma_{\mu\nu}(x)} + \delta_{\rm D}(v){\cal A}^a_\mu (x)\frac{\delta}{\delta {\cal A}^a_\mu(x)}\\
& \hspace{2.1cm} + \delta_{\rm D}(v){\cal C}^\mu (x)\frac{\delta}{\delta {\cal C}^\mu(x)} + \delta_{\rm D}(v)\bar{\cal C}_\mu (x)\frac{\delta}{\delta \bar{\cal C}_\mu(x)}\\
&\hspace{2.1cm} + \delta_{\rm D}(v)\Sigma^a (x)\frac{\delta}{\delta \Sigma^a(x)} + \delta_{\rm D}(v)\bar{\Sigma}^a (x)\frac{\delta}{\delta \bar{\Sigma}^a(x)}\bigg)
\end{aligned}\label{ward-diff} 
\end{equation}

\noindent and Yang-Mills gauge transformations:
\begin{equation}
\begin{aligned}
{\cal W}_{\rm YM}(\lambda) &\equiv -\int\:{\rm d}^d x\:\Big(\delta_{\rm YM}(\lambda)\gamma_{\mu\nu} (x)\frac{\delta}{\delta \gamma_{\mu\nu}(x)} + \delta_{\rm YM}(\lambda){\cal A}^a_\mu (x)\frac{\delta}{\delta {\cal A}^a_\mu(x)}\\
& \hspace{2.0cm} + \delta_{\rm YM}(\lambda){\cal C}^\mu (x)\frac{\delta}{\delta {\cal C}^\mu(x)} + \delta_{\rm YM}(\lambda)\bar{\cal C}_\mu (x)\frac{\delta}{\delta \bar{\cal C}_\mu(x)}\\
& \hspace{2.0cm} + \delta_{\rm YM}(\lambda)\Sigma^a (x)\frac{\delta}{\delta \Sigma^a(x)} + \delta_{\rm YM}(\lambda)\bar{\Sigma}^a (x)\frac{\delta}{\delta \bar{\Sigma}^a(x)}\Big)
\end{aligned}\label{ward-su}
\end{equation}

\noindent In these integrals the measure factor $\sqrt{{\rm det}(\gamma_{\mu\nu})}$ cancels against a similar one which would render the functional derivatives tensorial. We define the set of all invariant functionals by
\begin{eqnarray}\label{inv-set}
{\cal F}_{\rm inv} \equiv \{F\:|\:{\cal W}_{\rm D}(v) F = 0 \:\wedge\: {\cal W}_{\rm YM}(\lambda) F = 0 ~ \forall ~ v^\mu,\:\lambda^a\}
\end{eqnarray}

\noindent Computing the algebra of these operators leads to 
\begin{eqnarray}
\label{ward-d-d}[{\cal W}_{\rm D}(v_1), {\cal W}_{\rm D}(v_2)] &=& {\cal W}_{\rm D}([v_1, v_2])\\
\label{ward-l-l}[{\cal W}_{\rm YM}(\lambda_1), {\cal W}_{\rm YM}(\lambda_2)] &=& {\cal W}_{\textrm{YM}}(f \lambda_1\lambda_2)\\
\label{ward-d-l}[{\cal W}_{\rm D}(v), {\cal W}_{\rm YM}(\lambda)] &=& {\cal W}_{\rm YM}({\cal L}_v \lambda)
\end{eqnarray}

\noindent where $[v_1, v_2]$ denotes the Lie bracket and $(f \lambda_1\lambda_2)^a \equiv f^{abc}\lambda^b_1\lambda^c_2$\:. This algebra implies that the total group of gauge transformations, ${\bf G}$, has the structure of a semi-direct product of the spacetime diffeomorphisms ${\sf Diff}$ and the local Yang-Mills transformations ${\sf SU(N)}_{\rm loc}$, with the latter forming the invariant subalgebra: ${\bf G} = {\sf Diff} \ltimes {\sf SU(N)}_{\rm loc}$. Whereas the first two relations (\ref{ward-d-d}), (\ref{ward-l-l}) represent the well-known composition laws of diffeomorphisms and Yang-Mills gauge transformations, the third relation (\ref{ward-d-l}) lies at the heart of our problem: diffeomorphisms and local gauge transformations do not commute. Instead, they close on the Lie derivative of the gauge parameter. In particular, this implies that diffeomorphisms do not map $SU(N)$ tensors onto $SU(N)$ tensors. 

\subsection{Modified diffeomorphisms}
What is called for is an $SU(N)$ covariantization of the ordinary Lie derivative. This is tantamount to a different parametrization of ${\bf G}$ that makes the mixed commutator vanish. This can be achieved by defining {\it new diffeomorphisms} which include a ${\sf SU(N)}_{\rm loc}$ transformation with parameter $\lambda^a= \mathcal{A}^a_\mu v^\mu$ \cite{jackiw}:
\begin{eqnarray}\label{new-diff}
\widetilde{{\cal W}_{\rm D}}(v) \equiv {\cal W}_{\rm D}(v) + {\cal W}_{\rm YM} ({\cal A}\cdot v) 
\end{eqnarray}

\noindent Loosely speaking, this amounts to shifting a certain $v$-dependent part of ${\sf SU(N)}_{\rm loc}$ into the diffeomorphism sector. The notion of an ``invariant functional'' remains unchanged:
\begin{eqnarray}\label{inv-set-new}
{\cal F}_{\rm inv} = \{F\:|\:\widetilde{{\cal W}_{\rm D}}(v) F = 0 \:\wedge\: {\cal W}_{\rm YM}(\lambda) F = 0 ~ \forall ~ v^\mu,\:\lambda^a\} 
\end{eqnarray} 

\noindent The algebra relations receive extra contributions now since the Ward operators act on the field dependent parameters of the transformations as well. This leads to an algebra of the desired form (with $(v_1 v_2 \cdot F)^a\equiv v_1^\mu v_2^\nu F_{\mu\nu}^a$):
\begin{eqnarray}
\label{ward-d-d-new}[\widetilde{{\cal W}_{\rm D}}(v_1), \widetilde{{\cal W}_{\rm D}}(v_2)] &=& \widetilde{{\cal W}_{\rm D}}([v_1, v_2]) - {\cal W}_{\rm YM}(v_1 v_2\cdot F)\\
\label{ward-l-l-new}[{\cal W}_{\rm YM}(\lambda_1), {\cal W}_{\rm YM}(\lambda_2)] &=& {\cal W}_{\rm YM}(f \lambda_1\lambda_2)\\
\label{ward-d-l-new}[\widetilde{{\cal W}_{\rm D}}(v), {\cal W}_{\rm YM}(\lambda)] &=& 0
\end{eqnarray}

If we now split the dynamical fields into fluctuations and background configurations, we have to decide whether the field dependent transformation parameter in (\ref{new-diff}) should contain the full or the background gauge field only. Since, as already mentioned, the metric and gauge field fluctuations do not enter the final form of $S_{\rm gh}^{(2)}$ anyhow, we may safely opt for the latter already at this point:

\begin{equation}
\widetilde{\widetilde{{\cal W}_{\rm D}}}\!\!\!\raisebox{0.15cm}{{\scriptsize B,G}}(v)={\cal W}^{\rm B,G}_{\rm D}(v)+{\cal W}^{\rm B,G}_{\textrm{YM}}(\bar A \cdot v)
\end{equation}

According to the usual distinction between true and background gauge transformations, we now have to consider two classes of Ward operators as well. In the background case, the algebraic relations simply carry over, so we have for the Ward operators generating background gauge transformations:
\begin{eqnarray}
\label{b-ward-d-d-new}[\widetilde{\widetilde{{\cal W}^{\rm B}_{\rm D}}}(v_1), \widetilde{\widetilde{{\cal W}^{\rm B}_{\rm D}}}(v_2)] &=& \widetilde{\widetilde{{\cal W}^{\rm B}_{\rm D}}}([v_1, v_2]) - {\cal W}^{\rm B}_{\rm YM}(v_1 v_2\cdot \bar{F})\\
\label{b-ward-l-l}[{\cal W}^{\rm B}_{\rm YM}(\lambda_1), {\cal W}^{\rm B}_{\rm YM}(\lambda_2)] &=& {\cal W}^{\rm B}_{\rm YM}(f \lambda_1\lambda_2)\\
\label{b-ward-d-l}[\widetilde{\widetilde{{\cal W}^{\rm B}_{\rm D}}}(v), {\cal W}^{\rm B}_{\rm YM}(\lambda)] &=& 0
\end{eqnarray}

As for the true gauge transformations, the merely {\it background} field dependent parameter $\lambda^a= \bar{A}^a_\mu v^\mu$ of the compensating ${\sf SU(N)}_{\rm loc}$ transformation is not subject to true gauge transformations. The algebra that generates these transformations can be easily computed therefore by taking advantage of the linearity of commutators. We obtain
\begin{eqnarray}
\label{g-ward-d-d-new}[\widetilde{\widetilde{{\cal W}^{\rm G}_{\rm D}}}(v_1), \widetilde{\widetilde{{\cal W}^{\rm G}_{\rm D}}}(v_2)] &=& \widetilde{\widetilde{{\cal W}^{\rm G}_{\rm D}}}([v_1, v_2]) + {\cal W}^{\rm G}_{\rm YM}(v_1 v_2\cdot \bar{F})\\
\label{g-ward-l-l}[{\cal W}^{\rm G}_{\rm YM}(\lambda_1), {\cal W}^{\rm G}_{\rm YM}(\lambda_2)] &=& {\cal W}^{\rm G}_{\rm YM}(f \lambda_1\lambda_2)\\
\label{g-ward-d-l}[\widetilde{\widetilde{{\cal W}^{\rm G}_{\rm D}}}(v), {\cal W}^{\rm G}_{\rm YM}(\lambda)] &=& {\cal W}^{\rm G}_{\rm YM}(v\cdot\bar{\nabla} \lambda)
\end{eqnarray}

\noindent By writing $\widetilde{\widetilde{W}}$ we distinguish these Ward operators from the modified one that was defined with respect to the undecomposed fields; in addition, their notation shall remind us of the fact that the gauge field entered their definition only via its background component $\bar{A}$.

Thus the actual theory space on which we can define an RG flow consists of the functionals $F[h,a,\xi,\bar\xi,\Upsilon,\bar\Upsilon;\bar g,\bar A]\equiv F[g,\bar g,A,\bar A,\xi,\bar \xi,\Upsilon,\bar \Upsilon]$ in
\begin{equation}
{\cal F}'_{\rm inv} = \{F\:|\:\widetilde{\widetilde{{\cal W^{\rm B}_{\rm D}}}}(v) F = 0 \:\wedge\: {\cal W}^{\rm B}_{\rm YM}(\lambda) F = 0 ~ \forall ~ v^\mu,\:\lambda^a\}
\end{equation}

Finally, we return to our starting point and compute the $\bar{\Upsilon}$-$\xi$ part of the ghost action, now with the original true diffeomorphism $\delta_{\rm D}^{\rm G}$ replaced by its modified counterpart, $\widetilde{\widetilde{\delta_{\rm D}^{\rm G}}}$: 
\begin{equation}
\begin{aligned}
\left(S_{\rm gh}\:[\xi, \bar{\Upsilon}; \bar{g}, \bar{A}]\right)_{\bar{\Upsilon}\xi} &= - \int\:{\rm d}^d x\:\sqrt{\bar{g}}\:\big[\bar{\Upsilon}^a \bar{{\cal D}}^{\nu\:ab} \big(\widetilde{\widetilde{\delta_{\rm D}^{\rm G}}}(\xi)a^b_\nu\big)\big]|_{a=0}\\ 
&= - \int\:{\rm d}^d x\:\sqrt{\bar{g}}\:\Big[\bar{\Upsilon}^a \bar{{\cal D}}^{\nu\:ab} \Big(\xi^\rho \partial_\rho \bar{A}^b_\nu + (\partial_\nu \xi^\rho) \bar{A}^b_\rho\\& \hspace{3cm} -\big(\partial_\nu(\bar{A}^b_\rho \xi^\rho) + f^{bcd}\bar{A}^b\bar{A}^d_\rho \xi^\rho\big)\Big)\Big]\\
&= - \int\:{\rm d}^d x\:\sqrt{\bar{g}}\:\Big[ - \bar{\Upsilon}^a \bar{{\cal D}}^{\nu\:ab} \bar{F}^b_{\nu\rho} \xi^\rho\Big]
\end{aligned}\label{ups-xi-geist-neu}
\end{equation}
\noindent This action is obviously invariant under ${\sf SU(N)}_{\rm loc}$ and background diffeomorphisms $\widetilde{\widetilde{\delta_{\rm D}^{\rm B}}}$ for background tensorial ghost expectation values.
\section{The Running Gauge Coupling}
In this section we explicitly evaluate the FRGE on a truncated theory space which is general enough to allow for an approximate determination of the beta function for the scale dependent Yang-Mills coupling $g_\textrm{YM}(k)$. Our truncation is given by the following ansatz:
\begin{equation}
\begin{split}
\Gamma_k[g,\bar g, A,\bar A,\xi,\bar \xi,\Upsilon,\bar \Upsilon]=\Gamma^{\textrm{EH}}_k[g]+\Gamma^{\textrm{YM}}_k[g,A]+\Gamma^{\textrm{gf}}_k[g-\bar g,A-\bar A;\bar g,\bar A]\\ +S_{\textrm{gh}}[g-\bar g,A-\bar A,\xi,\bar \xi,\Upsilon,\bar \Upsilon;\bar g,\bar A]\label{ansatz}
\end{split}
\end{equation}
Here
\begin{equation}
\Gamma^{\textrm{EH}}_k[g]=2 \kappa^2 Z_N(k)\int \ddx \sqrt{g}\,\left(-R(g)+2\bar \lambda(k)\right)
\end{equation}
is a $k$-dependent form of the Einstein-Hilbert action. The corresponding dimensionful running parameters are the cosmological constant $\bar \lambda(k)$ and Newton's constant $G(k)\equiv \hat G /Z_N(k)$ where $\hat G$ is a fixed reference value. Furthermore,
\begin{equation}
\Gamma^{\textrm{YM}}_k[g,A]=\frac{Z_F(k)}{4\:\hat{g}^2_{\textrm{YM}}} \int \ddx \sqrt{g}\, g^{\mu\rho}g^{\nu\sigma} F^a_{\mu\nu}F^a_{\rho\sigma}
\end{equation}
is the standard second-order Yang-Mills action, with a $k$-dependent prefactor $Z_F(k)$ though. Hence the (dimensionful, except in d=4) running gauge coupling is $\bar g_\textrm{YM}(k)=\hat{g}_\textrm{YM} Z_F(k)^{-1/2}$ with some constant $\hat{g}_\textrm{YM}$. Finally,
\begin{equation}
\Gamma^{\textrm{gf}}_k[g-\bar g,A-\bar A;\bar g,\bar A]=\int \ddx \sqrt{\bar g}\, \left( \frac{Z_N(k)}{2 \alpha_{\textrm{D}}} \bar g^{\mu\nu} \textrm{F}_\mu\textrm{F}_\nu +\frac{Z_F(k)}{2\alpha_{\textrm{YM}}} \textrm{G}^a\textrm{G}^a\right)
\end{equation}
implements the gauge fixing conditions for the diffeomorphisms, $\textrm{F}_\mu$, and the $SU(N)$ gauge transformations, $\textrm{G}^a$. As in \cite{mr} and \cite{nonabavact} we factored out the wave function renormalizations $Z_N$ and $Z_F$ from the gauge fixing parameters $\alpha_\textrm{D}$ and $\alpha_\textrm{YM}$, respectively. In principle the latter are still $k$-dependent but we shall neglect their running here. In fact, later on we set $\alpha_\textrm{D}=\alpha_\textrm{YM}=1$. Our choice for the gauge conditions complies with the requirements discussed in the previous section:
\begin{align}
\textrm{F}_\mu(\bar h;\bar g)&=\sqrt{2}\kappa\left(\delta_\mu^\beta \bar g^{\alpha \gamma}\bar D_\gamma - \frac{1}{2} \bar g^{\alpha\beta}\bar D_\mu\right)\bar h_{\alpha\beta}\\
\textrm{G}^a(\bar a;\bar g,\bar A)&=\hat{g}^{-1}\bar g^{\mu\nu}\bar{\mathcal D}_\mu \bar{a}^a_\nu
\end{align}
The resulting ghost action reads, with $a^a_\mu \neq 0$ and $h_{\mu\nu}\neq 0$ still, 
\begin{multline}\label{SGhost}
S_{\textrm{gh}}[h,a,\xi, \bar \xi, \Upsilon, \bar \Upsilon;\bar g,\bar A]=\\
-\int \ddx \sqrt{\bar{g}}\,\left(\sqrt{2} \bar \xi_\mu\left(\bar g^{\mu\rho}\bar g^{\sigma\lambda}\bar D_\lambda\left(g_{\rho\nu}D_\sigma+g_{\sigma\nu}D_\rho\right)-\bar g^{\rho\sigma}\bar g^{\mu\lambda}\bar D_\lambda g_{\sigma\nu}D_\rho\right)\xi^\nu \right. \\
 + \bar\Upsilon^a \bar g^{\mu\nu} \bar{\mathcal D}_\mu ( \bar F^a_{\rho\nu} \xi^{\rho} + \xi^\rho \partial_\rho a^a_\nu +(\partial_\nu \xi^\rho)a^a_\rho + f^{abc}\bar{A}^b_\rho \xi^\rho a^c_\nu) + \bar\Upsilon^a\left(\bar g^{\mu\rho}\delta^{ab}\bar{\mathcal D}_\mu \nabla_\rho\right)\Upsilon^b\Big)
\end{multline}

It can be checked that $S_{\textrm{gh}}$ of eq. (\ref{SGhost}) is invariant under background gauge transformations: ${\cal W}^{\rm B}_{\rm YM} S_{\rm gh}=0=\widetilde{\widetilde{{\cal W}^{\rm B}_{\rm D}}} S_{\rm gh}$. While this is true even for non-vanishing fluctuations $h$ and $a$, in the present calculation we shall need $S_{\rm gh}$ only for $h=0=a$.

At this point a remark concerning the expected reliability of this truncation ansatz might be in order. As for its gravitational part, all generalizations of the Einstein-Hilbert truncation explored during the past decade did not change the qualitative picture it gives rise to, at least close to the non-Gaussian fixed point. In the Yang-Mills sector we retained only the first monomial of a systematic derivative expansion. From the analysis in \cite{nonabavact} without gravity we know that this truncation is not only sufficient to reproduce one-loop perturbation theory exactly, but even approximates the two-loop result for the beta-function with a small error of a few percent. Therefore we may expect that this truncation, too, is perfectly sufficient as long as $k$ is sufficiently large (well above dynamically generated ``confinement'' scales, say).

When we insert the truncation ansatz (\ref{ansatz}) into the exact FRGE (\ref{FRGE}) the supertrace decomposes into a ``bosonic'' and a ghost contribution:
\begin{align}
\partial_t\Gamma_k &= \frac 1 2 \,\textrm{Tr}\left[\frac{\partial_t \mathcal{R}_k\big(\mathcal{Z}^{-1}_k\breve{\Gamma}^{(2)}_k\big)}{\breve{\Gamma}^{(2)}_k+\mathcal{R}_k\big(\mathcal{Z}^{-1}_k\breve{\Gamma}^{(2)}_k\big)}\right]-\textrm{Tr}\left[\frac{\partial_t \mathcal{R}^{\textrm{gh}}_k\big(\mathcal{Z}^{-1}_{\textrm{gh}} S^{(2)}_{\textrm{gh}}\big)}{S^{(2)}_{\textrm{gh}}+\mathcal{R}^{\textrm{gh}}_k\big(\mathcal{Z}^{-1}_{\textrm{gh}}S^{(2)}_{\textrm{gh}}\big)}\right]\label{ERGE}
\end{align}
Here $\breve\Gamma_k\equiv \Gamma^\textrm{EH}_k+\Gamma^{\textrm{YM}}_k+\Gamma^{\textrm{gf}}_k$ is the bosonic part of the action and $\breve{\Gamma}^{(2)}_k$ is its Hessian. The coarse graining operators in (\ref{ERGE}) have the structure
\begin{equation}
\mathcal{R}_k(x) =\mathcal{Z}_k k^2 R^{(0)}(x/k^2)\qquad \mathcal{R}^{\textrm{gh}}_k(x) = \mathcal{Z}^{\textrm{gh}}_k k^2 R^{(0)}(x/k^2)
\end{equation}
where $R^{(0)}(y)$ is a ``shape function'' continuously interpolating between $R^{(0)}(0)=1$ and $\lim\limits_{y \rightarrow \infty}{ R^{(0)}(y)}=0$. The constants $\mathcal Z_k$ and $\mathcal Z^{\textrm{gh}}_k$ are matrices in field space. They will be adjusted in such a way that if in $\Gamma^{(2)}_k$ a certain mode has the inverse propagator $\zeta_k p^2$ it becomes $\zeta_k\left(p^2 +k^2 R^{(0)}\right)$ when we add $\mathcal R_k$ to $\Gamma^{(2)}_k$. As we shall see, this requirement is met if $\mathcal Z_k$ and $\mathcal Z^{\textrm{gh}}_k$ have the following block structure in $(\bar h,\bar a,\xi,\bar\xi,\Upsilon,\bar\Upsilon)$-space:
\begin{equation}
\begin{aligned}
\Big[\left(\mathcal{Z}_k\right)_{\bar h\bar h}\Big]^{\mu \nu}_{\ \ \rho\sigma} &=\frac{Z_N(k) \kappa^2}{2}(\delta^\mu_\rho \delta^\nu_\sigma+\delta^\nu_\rho \delta^\mu_\sigma-\bar g^{\mu\nu} \bar g_{\rho\sigma}) \\
\Big[\left(\mathcal Z^{\textrm{gh}}_k\right)_{\bar \xi \xi}\Big]^\mu_{\ \nu}&=\sqrt{2}\; \delta^\mu_\nu
\end{aligned}
\
\begin{aligned}
\Big[\left(\mathcal{Z}_k\right)_{\bar a\bar a}\Big]^{a \mu b}_{\ \ \ \nu} &= \frac{Z_F(k)}{\hat{g}^2}\; \delta^{ab} \delta^\mu_{\nu}\\
\Big[\left(\mathcal Z^{\textrm{gh}}_k\right)_{\bar \Upsilon \Upsilon}\Big]^{ab}&=\delta^{ab}
\end{aligned}
\label{Zfactors}
\end{equation}
Note that $\mathcal Z^{\textrm{gh}}_k$ is actually $k$-independent.

In setting up eq. (\ref{ERGE}) we followed ref. \cite{nonabavact} and opted for the complete Hessian operator $\Gamma^{(2)}_k$ to play the r\^{o}le of $\Delta$.\footnote{Recently this kind of cutoff has been referred to as ``spectrally adjusted'' \cite{litimspec,giesymfp,ymrev} or as of ``type III'' \cite{codello}.} More precisely, we set $\Delta= \mathcal Z^{-1}_k\breve{\Gamma}^{(2)}_k$ and $\Delta=\mathcal Z_{\textrm{gh}}^{-1} S^{(2)}_\textrm{gh}$ in the $(\bar h,\bar a)$- and the ghost-sectors, respectively. The multiplication by the inverse $\mathcal Z$ matrices brings $\Delta$ closer to an ordinary (covariant) Laplacian; symbolically, if $\Gamma^{(2)}_k= - \zeta_k \partial^2+ \cdots, \ \mathcal Z_k =\zeta_k$, we employ $\Delta=-\partial^2+\cdots$ rather than $\Delta=-\zeta_k \partial^2+\cdots$.

The most complicated ingredient needed in order to evaluate the traces in the FRGE is the Hessian of the bosonic action $\breve{\Gamma}^{(2)}_k$, i.\,e. the matrix of its second functional derivatives with respect to the dynamical fields $(\bar h, \bar a)$, or equivalently $(g, A)$, at fixed backgrounds $(\bar g, \bar A)$. This Hessian is most transparently displayed by means of the associated quadratic form $\Gamma^{\textrm{quad}}_k$ which appears in the expansion
\begin{equation}
\breve{\Gamma}_k[\bar g+\bar h,\bar A+ \bar a,\bar g,\bar A]= \breve{\Gamma}_k[\bar g, \bar A,\bar g,\bar A]+O(\bar h,\bar a)+\Gamma^{\textrm{quad}}_k[\bar h, \bar a;\bar g,\bar A]+{\cal O}(\{\bar h,\bar a\}^3)
\end{equation}
Explicitly, $\Gamma^{\textrm{quad}}_k$ is the sum of the following terms which reflect the block structure of $\breve{\Gamma}^{(2)}_k$ in $(\bar h,\bar a)$-space:
\begin{align}
\left(\Gamma^{\textrm{quad}}_k\right)_{\bar h \bar h}&=Z_N \kappa^2 \int \ddx \sqrt{\bar g}\,\bar h_{\chi\xi}\bigg(\left(U^{\chi\xi}_{\quad\eta\zeta}-\bar g^{\rho \sigma}\bar D_\rho \bar D_\sigma K^{\chi\xi}_{\quad\eta\zeta}\right)+\nonumber \\&\qquad + \left(1-\frac{1}{\alpha_{\textrm{D}}}\right)L^{\rho\sigma\chi\xi}_{\quad\quad\eta\zeta}\bar D_\rho\bar D_\sigma+ \frac{Z_F}{2 Z_N \kappa^2} N^{\mu\nu\rho\sigma\chi\xi}_{\qquad\quad\eta\zeta}\frac 1 4 \bar{F}^a_{\mu\nu}\bar{F}^a_{\rho\sigma}\bigg)\bar h^{\eta\zeta}\\
\left(\Gamma^{\textrm{quad}}_k\right)_{\bar a \bar a}&=\frac{Z_F}{2\,\hat{g}^2_{\textrm{YM}}}\int \ddx \sqrt{\bar g}\, \bar a^a_\xi\bigg(-\delta^{ab}\delta^{\xi}_{\eta}\bar g^{\rho\mu}\bar{\mathcal D}_\rho \bar{\mathcal D}_\mu +2 \bar g^{\xi\rho} f^{abc}\bar{F}^c_{\rho\eta}+\delta^{ab}\bar g^{\xi\rho}\bar R_{\rho\eta}+\nonumber \\ &\qquad +\left(1-\frac{1}{\alpha_{\textrm{YM}}}\right) \delta^{ab} \bar g^{\xi\rho}\bar{\mathcal D}_\rho \bar{\mathcal D}_\eta\bigg)\bar a^{b\eta}\\
\left(\Gamma^{\textrm{quad}}_k\right)_{\bar h \bar a}&=\frac{Z_F}{2\,\hat{g}^2_{\textrm{YM}}}\int \ddx \sqrt{\bar g}\,\bar h_{\eta\zeta}\left(\left(\frac 1 2 \delta_\xi^{\sigma}\bar g^{\eta\zeta}\bar g^{\mu\rho} +\delta_\xi^{\eta}\bar g^{\zeta\rho}\bar g^{\sigma\mu}+\delta_\xi^{\rho}\bar g^{\sigma\zeta}\bar g^{\mu\eta}\right)\bar F^a_{\rho\sigma}\bar{\mathcal D}_\mu\right)\bar a^{a\xi}\\
\left(\Gamma^{\textrm{quad}}_k\right)_{\bar a \bar h}&=\left(\Gamma^{\textrm{quad}}\right)_{\bar h \bar a}
\end{align}
The above quadratic functionals contain the kernels
\begin{align}
K^{\chi\xi}_{\quad\eta\zeta}&=\frac{1}{4}\left(\delta^\chi_\eta\delta^\xi_\zeta+\delta^\xi_\eta\delta^\chi_\zeta-\bar g^{\chi\xi}\bar g_{\eta\zeta}\right)\\
U^{\chi\xi}_{\quad\eta\zeta}&=\frac{1}{4}\left(\delta^\chi_\eta\delta^\xi_\zeta+\delta^\xi_\eta\delta^\chi_\zeta-\bar g^{\chi\xi}\bar g_{\eta\zeta}\right)\left(\bar R -2\bar\lambda\right)+\bar g^{\chi\xi}\bar R_{\eta\zeta}-\delta^\chi_\eta\bar R^\xi_{\;\zeta}-\bar R_{\zeta\ \eta}^{\ \chi\ \xi}\\
L^{\rho\sigma\chi\xi}_{\quad\quad\eta\zeta}&=\left(\frac 1 4 \bar g^{\chi\xi}\bar g^{\rho\sigma}\bar g_{\eta\zeta}-\frac 1 2 \delta^\rho_\eta\delta^\sigma_\zeta\bar g^{\chi\xi}-\frac 1 2 \bar g^{\chi\rho}\bar g^{\xi\sigma}\bar g_{\eta\zeta}+\delta^\chi_\eta\delta^\sigma_\zeta\bar g^{\xi\rho}\right)\\
N^{\mu\nu\rho\sigma\chi\xi}_{\qquad\quad\eta\zeta}&=\frac 1 2\left(\frac 1 2 \bar g^{\chi\xi}\bar g_{\eta\zeta}-\delta^\chi_\eta\delta^\xi_\zeta\right)\bar g^{\mu\rho}\bar g^{\nu\sigma}+\nonumber \\ &\qquad+2\left(\delta^\mu_\eta\delta^\rho_\zeta\bar g^{\nu\chi}\bar g^{\sigma\xi}-\delta^\mu_\eta\delta^\rho_\zeta\bar g^{\chi\xi}\bar g^{\nu\sigma}+2\delta^\xi_\eta\delta^\rho_\zeta\bar g^{\mu\chi}\bar g^{\sigma\nu}\right)
\end{align}

Using these formulae it can be checked that the $\mathcal Z$-factors (\ref{Zfactors}) are correctly chosen. Since we are not going to extract any ``extra'' background field dependence \cite{elisa2} we may set $\bar g_{\mu\nu}=g_{\mu\nu}$ and $\bar A^a_\mu=A^a_\mu$ after having found the Hessian.

The truncation contains three running couplings, $\bar{g}_\textrm{YM}(k)$, $G(k)$ and $\bar{\lambda}(k)$. Their beta functions can be found from the FRGE (\ref{ERGE}) by ``projecting out'' the corresponding invariants in the derivative expansion of the traces and equating them to the corresponding field monomials on the LHS of the flow equation. The resulting system of differential equations becomes autonomous if we employ the dimensionless counterparts of $\bar g_\textrm{YM}$, $G$ and $\bar\lambda$ respectively:
\begin{equation}
\gYM^2(k)\equiv k^{d-4} Z^{-1}_F(k) \hat{g}_\textrm{YM}^2,\quad g(k)\equiv\frac{k^{d-2}}{32 \pi Z_N(k) \kappa^2}=k^{d-2} G(k),\quad \lambda(k)\equiv k^{-2}\bar\lambda(k)
\end{equation}
In terms of these variables the three coupled RG equations have the structure
\begin{equation}
\begin{aligned}
\partial_t g_\textrm{YM}^2 &=\beta_\textrm{YM}\equiv\left(d-4+\eta_F\right)g_\textrm{YM}^2\\
\partial_t g &=\beta_g \equiv\left(d-2+\eta_N\right)g\\
\partial_t \lambda &=\beta_\lambda
\end{aligned}
\label{SDE}
\end{equation}
Here we introduced the anomalous dimensions related to the Yang-Mills and the gravitational field, respectively:
\begin{equation}
\eta_{F}=-\partial_t \ln Z_F,\qquad \eta_{N}=-\partial_t \ln Z_N
\end{equation}

In this paper we are only interested in the gravitationally corrected Yang-Mills beta function $\beta_\textrm{YM}$. Therefore it is sufficient to extract the $F_{\mu\nu}^2$-term from the derivative expansion of the traces. For identifying this monomial and reading off its prefactor we may insert any metric. We shall employ the most convenient choice, $g_{\mu\nu}=\bar g_{\mu\nu}=\delta_{\mu\nu}$. Furthermore, we set $\alpha_\textrm{D}=\alpha_\textrm{YM}=1$ from now on. The remaining calculation is in principle straightforward, but rather lengthy. One has to expand the traces up to terms with two fields $A^a_\mu(x)$ and two derivatives acting on them. Because of the built-in background gauge invariance those terms should combine to $F_{\mu\nu}^a F^{a \mu\nu}$. As a check we verified that this indeed happens.

Let us now discuss the result. Here we specialize for $d=4$ spacetime dimensions; for general $d$ the reader is referred to the Appendix. We present three different formulae for $\eta_F$; they differ with respect to the degree of ``RG improvement'' they take into account.

To start with, we ``switch off'' all RG improvements. This means that we discard all terms in $\partial_t\mathcal R_k$ on the RHS of the flow equation where $\partial_t$ hits either a $\mathcal Z_k$-factor or the $\Gamma^{(2)}_k$ in the argument of $\mathcal R_k$. In this way the evaluation of the FRGE amounts to a one-loop calculation, with a non-standard regulator though. We find
\begin{equation}
\eta_F=-\frac 6 \pi \,g\, \Phi^1_1(0)-\frac{11}{24\pi^2} N \gYM^2
\end{equation}
so that
\begin{equation}
\partial_t \gYM^2 =-\frac 6 \pi \,g\, \gYM^2\, \Phi^1_1(0)-\frac{11}{24\pi^2} N \gYM^4\label{4d1loop}
\end{equation}
Here $\Phi^1_1(0)$ is one of the integrals which were encountered in the pure gravity calculation \cite{mr} already:
\begin{equation}
\begin{aligned}
\Phi^p_n(w)&=\frac{1}{\Gamma(n)}\int_0^\infty\dz z^{n-1}\frac{R^{(0)}(z)-z R^{(0)}{}'(z)}{[z+R^{(0)}(z)+w]^p}\qquad & (n&>0)\\
\Phi^p_0(w)&=(1+w)^{-p} & (n&=0)
\end{aligned}
\end{equation}
The second contribution on the RHS of (\ref{4d1loop}) is the familiar ``asymptotic freedom'' term due to the self-interaction of the gauge bosons, while the first one, due to the virtual gravitons, is new.

Several comments are in order here.

\noindent{\bf(1)} The gravitational correction is manifestly cutoff scheme dependent, i.\,e. it depends, via $\Phi^1_1(0)$, on the shape function $R^{(0)}$. However, for any admissable choice of $R^{(0)}$ the constant $\Phi^1_1(0)$ is positive. As a result, the gravity term has a qualitatively similar impact on $\gYM(k)$ as the gauge boson loops, namely to drive $\gYM(k)$ smaller at larger $k$. It tends to speed up the approach of asymptotic freedom.

For the exponential cutoff $R^{(0)}(y)=y/(e^y-1)$, for instance, one finds $\Phi^1_1(0)=\pi^2/6$, while the ``optimized'' one \cite{opt}, $R^{(0)}(y)=(1-y)\Theta(1-y)$, yields $\Phi^1_1(0)=1$.

\noindent{\bf(2)} The gravitational correction, in perturbative language, originates from a quadratic divergence or, in FRGE language, a quadratic running with $k$. For this reason its scheme dependence is by no means surprising or alarming. Rather, it is the usual situation which is always encountered when the effective average action is applied to matter theories with a quadratic running of parameters, masses, say. However, one should note that the couplings in $\Gamma_k$ as such are not observable or ``physical'' quantities. The latter must be $R^{(0)}$-independent. This independence comes about by a compensation of the scheme dependence among different running couplings. (In truncations this compensation might not be perfect.) In general there will be compensations between effective propagators and vertices, for instance. Analogous remarks apply to the gauge fixing dependence.

\noindent{\bf(3)} The beta function for $\gYM^2$ depends on all three couplings, $\gYM^2$, $g$ and $\lambda$. In the approximation of (\ref{4d1loop}) it happens to be independent of $\lambda$, but it does depend on $g(k)\equiv k^2 G(k)$, the dimensionless Newton constant. Hence the differential equation for $\gYM$ cannot be solved in isolation. In principle the full system (\ref{SDE}) should be considered, and this would include the backreaction of the matter fields on the running of the gravitational parameters $g$ and $\lambda$. We shall not study this backreaction in the present paper. Instead, let us assume that the complete RG trajectory $k\mapsto(\gYM(k),g(k),\lambda(k))$ admits a classical regime \cite{h3} in which Newton's constant does not run appreciably so that we may approximate 
\begin{equation}
G(k)\approx G_0=\textrm{const},\quad g(k)=G_0 k^2\label{Gassumption}
\end{equation}
This approximation, implicitly, has been made in all perturbative studies \cite{ymg-robwil,ymg-robthesis,ymg-piet,ymg-toms,ymg-ebert,ymg-tangwu,ymg-toms2}. With (\ref{Gassumption}), for an abelian field $(N=0)$, say,
\begin{equation}
\partial_t \gYM^2 =-\frac 6 \pi\, \Phi^1_1(0) \,G_0\,k^2\, \gYM^2\label{abelianresult}
\end{equation}
Incidentally this beta function has the same general structure as the result by Robinson and Wilczek \cite{ymg-robwil}; it is proportional to $G_0\gYM^2$ and depends explicitly on the energy scale $k$. Its $k^2$-dependence indicates that the underlying quantum effect is related to a quadratic divergence. 

Eq. (\ref{abelianresult}) is easily solved: $\gYM^2(k)=\gYM^2(0)\cdot \exp \left(-\omega_\textrm{YM}(k/m_\textrm{Pl})^2 \right)$. Here $\omega_\textrm{YM}\equiv 3 \Phi^1_1(0)/\pi$ and $m_\textrm{Pl}\equiv G_0^{-1/2}$ is the (ordinary, constant) Planck mass. To first order in the $k/m_\textrm{Pl}$-expansion we get
\begin{equation}
\gYM^2(k)=\gYM^2(0)\left[1-\omega_\textrm{YM} (k/m_\textrm{Pl})^2+O(k^4/m_\textrm{Pl}^4)\right]\label{kmexpansion}
\end{equation}
We note that to leading order Newton's constant itself \cite{mr} has an analogous scale dependence, including the sign of the correction: $G(k)=G_0\left[1-\omega (k/m_\textrm{Pl})^2+\cdots\right]$.

\noindent{\bf(4)} In order to illustrate how the above result fits into the asymptotic safety picture of Quantum Einstein Gravity \cite{wein,oliverbook,livrev} we consider a free Maxwell field again. It is known that, in the Einstein-Hilbert truncation, the RG flow of the average action possesses a non-Gaussian fixed point for the two gravitational couplings, $(g^*,\lambda^*)$, both in pure gravity \cite{mr,souma,oliver1} and in presence of a free Maxwell field \cite{perper1}. At this fixed point the \emph{dimensionless} Newton constant equals a positive constant, $g(k)=g^*$, while the dimensionful one runs to zero quadratically: $G(k)=g^*/k^2\rightarrow 0$ for $k\rightarrow \infty$. In this regime,
\begin{equation}
\partial_t \gYM^2 =-\frac 6 \pi\, \Phi^1_1(0) \,g^*\, \gYM^2
\end{equation}
The solution to this equation reads
\begin{equation}
\gYM^2(k)\propto k^{-\Theta_{\textrm{YM}}},\quad\Theta_{\textrm{YM}}=\frac 6 \pi \,g^*\,\Phi^1_1(0)\label{critexp}
\end{equation}
At the fixed point the gauge coupling approaches zero according to a power law with a critical exponent $\Theta_\textrm{YM}$, a positive number of order unity.\footnote{One cannot easily extract the precise numerical value of $\Theta_\textrm{YM}$ from existing calculations since the determination of $g^*$ in the Einstein-Maxwell system in ref. \cite{perper1} employs a cutoff different from the present one.} Thus the total system has a non-trivial fixed point of the form $(\gYM^*=0,g^*>0,\lambda^*>0)$. Obviously the approach of $\gYM=0$ is much faster than without gravity where $\gYM(k)\propto 1 / \ln(k).$ Note that $\gYM$ is a \emph{relevant} parameter, it grows when $k$ is lowered, hence it contributes one unit to the dimensionality of the fixed point's UV critical manifold.

Next we present the results for $\eta_F$ with the RG improvements included. In a first step we retain only the terms which arise when $\partial_t$ hits the $\mathcal Z_k$-factors in $\mathcal R_k$. Those terms are proportional to $\eta_F$ and $\eta_N$, respectively. As now $\eta_F$ appears also on the RHS of the RG equation we obtain an implicit equation for it. Its solution reads
\begin{equation}
\eta_F=\frac{-\frac{6}{\pi}\,g\,\Phi^1_1(0)-\frac{11}{24\pi^2} N \gYM^2-\frac{2}{\pi}\,\eta_N\, \lambda\, g}{1-\frac{3}{\pi}\, g\, \tilde{\Phi}^1_1(0)-\frac{5}{24 \pi^2} N \gYM^2 -\frac{2}{\pi}\,\lambda\, g}\label{4d1stimproved}
\end{equation}
In this approximation $\eta_F$ depends not only on Newton's but also on the cosmological constant. Eq. (\ref{4d1stimproved}) resums terms of arbitrary order both in $\gYM$ and $g$; it generalizes a known result \cite{nonabavact} for pure Yang-Mills theory. If we go on and include also the terms coming from the scale derivative of $\Gamma^{(2)}_k$ in the argument of $\mathcal{R}_k$ we are led to
\begin{equation}
\eta_F=\frac{-\frac{6}{\pi}\,g\,\Phi^1_1-\frac{11}{24\pi^2} N \gYM^2+\frac{1}{\pi}\,\eta_N\,g\, \left(3\tilde{\tilde{\Phi}}^1_2-2\lambda-2\lambda^2\,\tilde{\tilde{\Phi}}^1_0\right)-\frac{4}{\pi}\, g\, \lambda (2\lambda+\beta_\lambda)\tilde{\tilde{\Phi}}^1_0}{1-\frac{3}{\pi}\, g\, \left(\tilde{\Phi}^1_1-\tilde{\tilde{\Phi}}^1_2\right)-\frac{5}{24 \pi^2} N \gYM^2 -\frac{2}{\pi}\,\lambda\, g-\frac{2}{\pi}\,\lambda^2\, g\,\tilde{\tilde{\Phi}}^1_0}\label{4d2ndimproved}
\end{equation}
The eqs. (\ref{4d1stimproved}) and (\ref{4d2ndimproved}) contain the following integrals involving $R^{(0)}$:
\begin{equation}
\begin{aligned}
\tilde{\Phi}^p_n(w)&=\frac{1}{\Gamma(n)}\int_0^\infty\dz z^{n-1}\frac{R^{(0)}(z)}{[z+R^{(0)}(z)+w]^p} \qquad & (n&>0)\\
\tilde{\Phi}^p_0(w)&=(1+w)^{-p} & (n&=0)\\
\tilde{\tilde{\Phi}}^p_n(w)&=\frac{1}{\Gamma(n)}\int_0^\infty\dz z^{n-1}\frac{R^{(0)}{}'(z)}{[z+R^{(0)}(z)+w]^p}\qquad & (n&>0)\\
\tilde{\tilde{\Phi}}^p_0(w)&=\frac{R^{(0)}{}'(0)}{(1+w)^p}\qquad & (n&=0)
\end{aligned}
\end{equation}
The threshold functions $\tilde{\Phi}^p_n(w)$ appeared already in the Einstein-Hilbert truncation of pure gravity \cite{mr}. In eq. (\ref{4d2ndimproved}) we abbreviated $\tilde{\Phi}^p_n\equiv \tilde{\Phi}^p_n(0)$ and $\tilde{\tilde{\Phi}}^p_n\equiv \tilde{\tilde{\Phi}}^p_n(0)$, respectively. What is new in (\ref{4d2ndimproved}) is the occurrence of the complete beta function of the cosmological constant, $\beta_\lambda$. It originates from the differentiation of the $\bar\lambda$-term contained in $\Gamma^{(2)}_k$.

\section{Discussion and Conclusion}
In the recent literature on the gravitational corrections to the Yang-Mills beta function \cite{ymg-robwil,ymg-robthesis,ymg-piet,ymg-toms,ymg-ebert,ymg-tangwu,ymg-toms2} there has been a certain amount of confusion as some of the computations do get a non-zero result while others don't. However, we believe that different calculations have no reason to yield the same result unless they agree on virtually all details of the regularization and renormalization procedure. The quantum effects of interest are related to quadratic divergences (or a $k^2$-running), and so we should not expect the same high degree of universality as in the case of the familiar gauge boson contribution which is related to a logarithmic divergence.

In the present paper we computed the beta function for $\gYM(k)$, defined as a coefficient in the derivative expansion of the effective average action. This approach has two features which are essential here: First, it retains all quadratic divergences (as opposed to dimensional regularization, say), and second, by the background field technique, the regularization (the cutoff $\mathcal R_k$) preserves gauge invariance.\footnote{As in all traditional applications of the average action, $\Gamma_k$ is not gauge fixing independent, though, i.\,e. we do not use the Vilkovisky-DeWitt method.} In this setting, we do get a non-zero gravitational correction. This correction is scheme and gauge fixing dependent but, as we explained, this is by no means unexpected but rather the usual situation. When observable quantities are computed from $\Gamma_k$ the scheme and gauge fixing dependences will cancel among the different running couplings involved.

Among the perturbative calculations only the one by Tang and Wu \cite{ymg-tangwu} is directly comparable to ours. They employ a regulator which retains quadratic divergences and treats them in a gauge invariant manner. It is gratifying to see that they, too, get a non-zero gravitational correction which has the same structure as ours when we omit the RG improvements.

As a first application of our results we mention that, by a standard argument, knowledge about the $k$-dependence of wave function normalization constants such as $Z_F(k)$ can be used in order to deduce information about the related fully dressed propagator implied by $\Gamma\equiv\Gamma_{k=0}$. In the case at hand the running inverse propagator of the gauge field, on a flat background, has the form $Z_F(k)p^2\propto \gYM^{-2}(k)p^2$. At high momenta, if there is no other relevant physical cutoff scale but the momentum itself, the dressed propagator $D(p)$ obtains by setting $k=|p|$, whence $D(p)^{-1}\propto \gYM^{-2}(|p|)p^2$. For the example of eq. (\ref{kmexpansion}), for instance, this leads us to expect that the photon propagator gets modified by a $p^4$-term when $p$ approaches the Planck scale:
\begin{equation}
D(p)^{-1}=p^2+\omega_\textrm{YM}\: p^4/m_\textrm{Pl}^2+O(p^6/m_\textrm{Pl}^4)
\end{equation}
Likewise the fixed point running of (\ref{critexp}) implies the following behavior for $p^2 \rightarrow \infty$:
\begin{equation}
D(p)\propto 1/p^{2(1+\Theta_\textrm{YM}/2)} \label{propagator}
\end{equation}
As $\Theta_\textrm{YM}$ is positive the gauge field propagator falls off faster than $1/p^2$, thanks to the quantum gravity corrections. In fact, the same argument when applied to the graviton propagator leads to a $1/p^4$-behavior for $p^2 \rightarrow \infty$ \cite{oliver1}. The asymptotic propagator (\ref{propagator}) suggests that the quantum gravity corrections improve the finiteness properties of the matter field theory, and this precisely fits into the picture of asymptotic safety. It is also interesting to note that (\ref{propagator}) leads to a modified static electromagnetic potential $A_0(r)$ of a classical point charge.\footnote{Treating the source dynamically, also form factor effects need to be included \cite{reusch}} The 3-dimensional Fourier transform of (\ref{propagator}) yields the potential $A_0(r)\propto r^{\Theta_{\textrm{YM}}-1}$ which, if $\Theta_\textrm{YM}$ is large enough, could even be regular at $r=0$. This makes it obvious that the gravity induced running of the gauge coupling is closely related to the old problem of divergent self energies.

\appendix
\section{Appendix}
In this appendix we display the equation for $\eta_F$ in $d$ spacetime dimensions. With all RG improvements included, it assumes the form:
\begin{equation}
\begin{split}
&(4\pi)^{d/2}\eta_F=-32 \pi g \bigg[A \chi_{\frac{d}{2}-1,0}(-2\lambda)+B d^{-d/2} \chi_{\frac{d}{2}-1,0}(-2 \lambda d)\\[-0.1cm]
&\qquad+C (2\lambda)^{-1}\left(\chi_{\frac{d}{2},0}(-2\lambda)-\chi_{\frac{d}{2},0}(0)\right)\\[-0.1cm]
&\qquad+D\left\{d^{-d/2}\chi_{\frac{d}{2}+1,2}(-2\lambda d)-\chi_{\frac{d}{2}+1,2}(0)-\frac{d}{2}\left(d^{-d/2}\chi_{\frac{d}{2},1}(-2\lambda d)-\chi_{\frac{d}{2},1}(0)\right)\right\}\bigg]\\[-0.1cm]
&\quad-\gYM^2N\left(\frac{26-d}{3}\right)\chi_{\frac{d}{2}-2,0}(0)\\[-0.1cm]
&\quad +16 \pi \eta_N g \Bigg[A \tilde{\chi}_{\frac{d}{2}-1,0}(-2\lambda)+B d^{-d/2} \tilde{\chi}_{\frac{d}{2}-1,0}(-2 \lambda d)\\[-0.1cm]
&\qquad +C \left((2\lambda)^{-1}\tilde{\chi}_{\frac{d}{2},0}(-2\lambda)-(2\lambda)^{-2}\left(\tilde{\chi}_{\frac{d}{2}+1,0}(-2 \lambda)-\tilde{\chi}_{\frac{d}{2}+1,0}(0)\right)\right)\\[-0.1cm]
&\qquad -D\left\{\frac{d^{-d/2}}{d-1}\tilde{\chi}_{\frac{d}{2}+1,2}(-2\lambda d)+\frac{d^{-d/2+1}}{2}\tilde{\chi}_{\frac{d}{2},1}(-2\lambda d)-\frac{1}{d-1}\tilde{\chi}_{\frac{d}{2}+1,2}(0)\right\}\Bigg]\\[-0.1cm]
&\quad +16 \pi \eta_F g \left[C \left(-(2\lambda)^{-1}\tilde{\chi}_{\frac{d}{2},0}(0)+(2\lambda)^{-2}\left(\tilde{\chi}_{\frac{d}{2}+1,0}(-2 \lambda)-\tilde{\chi}_{\frac{d}{2}+1,0}(0)\right)\right)\right.\\[-0.1cm]
&\qquad \left.-D\left\{-\frac{d^{-d/2+1}}{d-1}\tilde{\chi}_{\frac{d}{2}+1,2}(-2\lambda d)-\frac{d}{2}\tilde{\chi}_{\frac{d}{2},1}(0)+\frac{d}{d-1}\tilde{\chi}_{\frac{d}{2}+1,2}(0)\right\}\right]\\[-0.1cm]
&\quad+\eta_F\gYM^2N\left(\frac{24-d}{6}\right)\tilde{\chi}_{\frac{d}{2}-2,0}(0)\\[-0.1cm]
&\quad-16\pi (\eta_F-\eta_N) g \bigg[A \tilde{\tilde{\chi}}_{\frac{d}{2},0}(-2\lambda)+B d^{-d/2} \tilde{\tilde{\chi}}_{\frac{d}{2},0}(-2 \lambda d)\\[-0.1cm]
&\qquad+C (2\lambda)^{-1}\left(\tilde{\tilde{\chi}}_{\frac{d}{2}+1,0}(-2\lambda)-\tilde{\tilde{\chi}}_{\frac{d}{2}+1,0}(0)\right)\\[-0.1cm]
&\qquad+D\frac{(d-2)}{2}\left\{\tilde{\tilde{\chi}}_{\frac{d}{2}+1,1}(0)-\tilde{\tilde{\chi}}_{\frac{d}{2}+1,1}(-2\lambda d)d^{-d/2}\right\}\bigg]\\[-0.1cm]
&\quad+32\pi g (2\lambda+\beta_\lambda)\Bigg[A \tilde{\tilde{\chi}}_{\frac{d}{2}-1,0}(-2\lambda)+B d^{-d/2+1} \tilde{\tilde{\chi}}_{\frac{d}{2}-1,0}(-2 \lambda d)\\[-0.1cm]
&\qquad +C \left((2\lambda)^{-1}\tilde{\tilde{\chi}}_{\frac{d}{2},0}(-2\lambda)-(2\lambda)^{-2}\left(\tilde{\tilde{\chi}}_{\frac{d}{2}+1,0}(-2\lambda)-\tilde{\tilde{\chi}}_{\frac{d}{2}+1,0}(0)\right)\right)\\[-0.1cm]
&\qquad -2D d\left\{\frac{d^{-d/2}}{d-1} \tilde{\tilde{\chi}}_{\frac{d}{2}+1,2}(-2\lambda d)-\frac{1}{d-1} \tilde{\tilde{\chi}}_{\frac{d}{2}+1,2}(0)+\frac{d}{2} d^{-d/2} \tilde{\tilde{\chi}}_{\frac{d}{2},1}(-2\lambda d)\right\}\Bigg]
\end{split}\label{result}
\end{equation}
Eq. (\ref{result}) contains the following functions of $d$:
\begin{equation}
\begin{aligned}
A&=\frac{(d+2)(d^2-9d+12)}{4d}\\[0.5cm]
C&=\frac{4(3d^2-10d+4)}{d(d-2)}
\end{aligned}
\qquad
\begin{aligned}
B&=\frac{(d-4)(d-6)}{2(d-2)}\\[0.5cm]
D&=\frac{4(d-4)^2}{d(d-2)^2(d-1)}
\end{aligned}
\end{equation}
Furthermore we introduced the following new threshold functions:
\begin{equation}
\begin{aligned}
\chi_{n,m}(w,v)&=\frac{1}{\Gamma(n-m)}\int_w^\infty\dz \frac{(z-w)^{n-1}}{(z+v)^m}\frac{R^{(0)}(z)-z R^{(0)}{}'(z)}{z+R^{(0)}(z)} \\
\tilde{\chi}_{n,m}(w,v)&=\frac{1}{\Gamma(n-m)}\int_w^\infty\dz \frac{(z-w)^{n-1}}{(z+v)^m}\frac{R^{(0)}(z)}{z+R^{(0)}(z)} \qquad\qquad\qquad (n>m\geq0)\\
\tilde{\tilde{\chi}}_{n,m}(w,v)&=\frac{1}{\Gamma(n-m)}\int_w^\infty\dz \frac{(z-w)^{n-1}}{(z+v)^m}\frac{R^{(0)}{}'(z)}{z+R^{(0)}(z)}
\end{aligned}
\end{equation}
When used in Eq. (\ref{result}) with one argument only, the second argument of the $\chi$ functions, $v$, is set to $v=-\frac{2\lambda d}{d-1}$ implicitly, for clarity purposes.

For $n \in \mathbb{N}$ and $m=0$ the $\chi$ integrals are independent of $v$ and can be expanded in $w$ into a \emph{finite} sum of $\Phi$ integrals,
\begin{equation}
\chi_{n,0}(w)=\sum_{k=0}^n \frac{(-w)^k}{k!}\Phi^1_{n-k}(0),
\end{equation}
and analogously for $\tilde{\chi}$ and $\tilde{\tilde{\chi}}$. For $d=4$ only $\chi$ integrals that are expandable in this way occur.

For $m>0$ a non-integrable singularity might occur in the integration interval (depending on the signs of $w$ and $v$). In this case the definition of the $\chi$ integrals is meant merely as an abbreviation, in the following sense. The integrands of the $\chi$ integrals embraced by curly brackets in (\ref{result}) should be summed first and only then be integrated over the intersection of their integration intervals. This removes the apparent singularity and we are left with a sum of finite integrals.

Eq. (\ref{result}) can be reduced to the corresponding 1-loop result by omitting from its RHS all terms containing $\eta_F$, $\eta_N$, or any $\tilde{\tilde{\chi}}$ integral. The latter arise from the differentiation of $\breve{\Gamma}^{(2)}_k$ in the arguments of $\mathcal R_k$.

\end{document}